\documentclass[twocolumn,showpacs]{revtex4}
\usepackage{graphics,epsf,bm}
\def\h4s{$^{\hspace*{0.1em}4}_\Sigma$He }
\def\bK{\mbox{\boldmath $K$}}
\def\bR{\mbox{\boldmath $R$}}
\def\bk{\mbox{\boldmath $k$}}
\def\bp{\mbox{\boldmath $p$}}
\def\br{\mbox{\boldmath $r$}}
\def\bs{\mbox{\boldmath $s$}}
\begin{document}
\draft
\title{Semiclassical Distorted Wave Model
Analysis of the $(\pi^-,K^+)$ $\Sigma$ Formation Inclusive Spectrum}
\author{M. Kohno,$^1$ Y. Fujiwara,$^2$ Y. Watanabe,$^3$
K. Ogata$^4$ and M. Kawai$^4$}
\affiliation{$^1$Physics Division, Kyushu Dental College,
Kitakyushu 803-8580, Japan\\
$^2$Department of Physics, Kyoto University,
Kyoto 606-8502, Japan\\
$^3$Department of Advanced Energy Engineering Science,
Kyushu University, Kasuga, Fukuoka 816-8580, Japan\\
$^4$Department of Physics, Kyushu University,
Fukuoka 812-8581, Japan}

\begin{abstract}
$(\pi^-,K^+)$ hyperon production inclusive spectra with $p_\pi =1.2$ GeV/c
measured at KEK on $^{12}$C and $^{28}$Si are analyzed by the
semiclassical distorted wave model. Single-particle wave functions of
the target nucleus are treated using Wigner transformation. This method is
able to account for the energy and angular dependences of the elementary
process in nuclear medium without introducing the factorization
approximation frequently employed.
Calculations of the $(\pi^+,K^+)$ $\Lambda$ formation process,
for which there is no free parameter since the $\Lambda$ s.p.
potential is known, demonstrate that the present model is useful to
describe inclusive spectra. It is shown that in order to account for the
experimental data of the $\Sigma^-$ formation spectra a repulsive
$\Sigma$-nucleus potential is necessary whose magnitude is not
so strong as around 100 MeV previously suggested.
\end{abstract}

\maketitle

\section{Introduction}
Various meson production reactions in nuclei are a rich source
of our understanding of hadronic interactions. In particular,
interactions involving hyperons have to be explored by
strangeness exchange reactions, since hyperons are absent
in ordinary nuclear systems. Naturally, the absence itself is
a consequence of the properties of strange hadrons.
The interaction between the lambda hyperon and the nucleon
is fairly well known, since the experimental data for $\Lambda$
hypernuclei has been accumulated in last more than 30 years.
The $\Sigma$-$N$ and $\Xi$-$N$ interactions, by contrast, have
not been well understood. Even the sign of the $\Sigma$
single-particle (s.p.) potential in nuclear medium, which
reflects basic properties of the $\Sigma$-$N$ interaction,
has not been established. Interactions among hyperons are
far less investigated, except for the $\Lambda \Lambda$ case.

In recent years, much experimental effort has been directed to
the study of strange baryons and baryonic resonances in
nuclear medium, using incident $\pi$, $K$ and $\gamma$ beams
with the energy of 1 $\sim$ 2 GeV.
The extraction of meaningful understanding of these baryon
properties from such reaction processes is not so simple, however.
For analyzing experimental data we need to take into account
various effects, such as the proper treatment of projectiles and
outgoing hadrons, the model description of elementary
processes in nuclei and the decent description of the target nucleus
and the residual (hyper) nucleus.
We also have to keep in mind the possibility of the change of
properties of the relevant hadrons themselves in nuclear
medium.

Since fully microscopic description is far from practical,
various approximations are commonly introduced to analyze the
experimental data. Then, it is important to employ a model
as simple and reliable as possible, bearing in mind that lack of
some proper treatment can easily lead to
misunderstanding of the basic hadron properties.

$\Sigma$ formation spectra in $(\pi, K)$ and $(K, \pi)$ reactions
with nuclei are not expected to have narrow peaks,
because of the strong $\Sigma N\rightarrow \Lambda N$ coupling.
In spite of this, however, the early $(K, \pi)$ experimental
spectra \cite{BERT} were interpreted as indicating an attractive
$\Sigma$ s.p. potential with the depth of
about 10 MeV \cite{KHW,DMG}.
The experimental discovery of \h4s \cite{HAYA,NAGA} has shown
that the $\Sigma$-$N$ interaction in the $T=1/2$ channel is
sufficiently attractive to support the bound state in this specific
nucleus, as discussed
by Harada \cite{HARA}. It has been recognized, however, that due to
the strong repulsion in the isospin $T=3/2$ channel, $\Sigma$ bound
states are unlikely to be observed in heavier nuclei. This conjecture was
supported by experimental results on targets of $^6$Li and $^9$Be
measured at BNL \cite{BART}. The analysis of  the $(K, \pi)$ spectra
on $^9$Be from BNL \cite{BNL} given by D\c{a}browski \cite{DAB} in
a plane wave impulse approximation method suggested that the
$\Sigma$ potential is repulsive of the order of 20 MeV.

The shift and the width of $\Sigma^-$ atomic states are another
source of the information on the $\Sigma$-nucleus potential.
Batty, Friedman and Gal \cite{BFG} reexamined the $\Sigma^-$
atomic data and concluded that the $\Sigma$ potential should be
attractive at the surface region but changes its sign to become repulsive
at the higher density region in a nucleus.

Theoretical studies for the two-body $\Sigma$-$N$ force have also
been inconclusive.
In the 1970s, the Nijmegen group started to construct hard-core
hyperon-nucleon potentials in a one-boson exchange model.
Parameter sets corresponding to two typical choices of the
SU(3) mixing angles were named as models D and F \cite{NIJDF}.
The $G$-matrix calculation by Yamamoto and Bando in Ref. \cite{YB1}
showed that the model D yields $-29.3$ MeV for the $\Sigma$ potential
in nuclear matter at the normal density ($k_F=1.35$ fm$^{-1}$) and
the model F repulsive 5.8 MeV, though the explicit numbers vary in a
different calculational scheme. The soft-core versions subsequently
constructed by the Nijmegen group \cite{NIJNS} tend to predict an
attractive $\Sigma$ s.p. potential in nuclear medium; $-27.1$ MeV
in Ref. \cite{YB1} and $-15.3$ MeV in the nuclear matter calculation
by Schulze {\it et al.}\cite{SCHU}.

A different approach using a non-relativistic SU(6) quark model has
been developed by the Kyoto-Niigata group \cite{FU96a,FU96b,FU01} to
obtain a unified description of octet baryon-baryon interactions.
In this model, the description of the short-ranged part of
baryon-baryon interactions basically provided by the resonating-group
method with the spin-flavor SU(6) quark model wave functions and the
one-gluon exchange Fermi-Breit interaction is supplemented by effective
meson-exchange potentials acting between quarks. This model has
little ambiguities in the hyperon-nucleon
sector after the nucleon-nucleon interaction is determined. $G$ matrix
calculations in the lowest order Brueckner theory \cite{KOH}
with the potential named as FSS \cite{FU96a,FU96b} show that the
$\Sigma$ s.p. potential in symmetric nuclear matter is repulsive of
the order of 20 MeV at normal density.
The repulsion due to a strongly repulsive character in the isospin
$T=\frac{3}{2}$ $^3S_1$ channel originating from quark Pauli effects
overcomes an attractive contribution in the $T=1/2$ $^3S_1$ channel
which is similar to that in the $\Lambda N$ case. The latest version of
this quark model potential, fss2 \cite{FU01}, gives a smaller
repulsion of about 8 MeV in symmetric nuclear matter. 

We briefly mention relativistic mean filed model description for the
hyperon sector. This model does not seem to have much predictive power, but
once parameters are determined to fit basic properties it has wide
applicability, for example, to a variety of neutron star matter calculations.
In early models including only $\sigma$ meson, the $\Sigma$ hyperon
is predicted to have the similar attractive potential to the $\Lambda$
in nuclear medium. Having recognized that the $\Sigma$ s.p. potential
may be repulsive in nuclei, the model was extended to include
$\sigma^*$ meson to account for that property. The repulsion of 30 MeV
has been tentatively used in literature \cite{MFGJ,SBG}, although this
specific number did not have a solid basis.

It is also noted that Kaiser \cite{KAIS} calculated the $\Sigma$
mean-field in symmetric nuclear matter in the framework of SU(3) chiral
perturbation theory and found a moderately repulsive potential, that is,
59 MeV for the real part and -21.5 MeV for the imaginary one at normal
density.

Recently, inclusive $(\pi^-,K^+)$ spectra corresponding to $\Sigma$
formation were measured at KEK \cite{KEK,SAHA} with better accuracy
than before, using the pion beam with the momentum of
$p_\pi = 1.2$ GeV/c on medium-to heavy nuclear targets. DWIA
analyses in Ref. \cite{KEK} for $^{28}$Si and similar analyses later
on other nuclear targets \cite{SAHA} gave a notable conclusion that
the $\Sigma$ potential is strongly repulsive, as large as 100 MeV.
Harada and Hirabayashi \cite{HH} showed in similar calculations with their
optimal Fermi-averaging for the elementary $t$-matrix that the $\Sigma$
potential is repulsive inside the nuclear surface, though the actual
strength varies with the imaginary part supposed.

The determination of the $\Sigma$-$N$ interaction is of fundamental
importance in the study of such problems as those of neutron star
matter and heavy ion collisions, because the baryonic component of
such hadronic matter, especially the hyperon admixture, is governed
by the basic baryon-baryon interactions. Considering the importance
of determining the $\Sigma$-$N$ interaction on the basis of
experimental data, it is desirable to analyze the KEK experiments
in a different and independent calculational scheme from those
in Refs. \cite{KEK, HH}. In this paper, we present a semiclassical method
for the DWIA approach and apply it to $(\pi^\pm,K^+)$ inclusive
spectra. The preparatory version of this approach was reported
in Ref. \cite{MK}. The semiclassical distorted wave (SCDW) model was
originally considered for describing intermediate energy nucleon
inelastic reactions on nuclei \cite{SCDW1}. Applications to various
$(p,p')$ and $(p,n)$ inclusive spectra \cite{SCDW2,SCDW3} have
demonstrated that the method is quantitatively reliable and thus the
applications to the wide range of nuclear reactions are promising.

In Sec. II, we show basic expressions of the semiclassical distorted
wave model for describing the $(\pi ,K)$ inclusive spectra. The formulation
using the Wigner transformation for the nuclear density matrix is explained.
Actual optical potential model parameters used for incident pions and outgoing
kaons are given in Sec. III. Then, numerical results for the $(\pi^\pm,K^+)$
spectra are presented in Sec. IV: first for the $\Lambda$ formation to
see the applicability of the SCDW model and next for the $\Sigma$ formation.
The latter case is the main concern of the present
paper to obtain more solid information about the strength of the $\Sigma$
single-particle potential than before. In this paper, we are concerned
with the spectra on light nuclei with $N=Z$, namely $^{28}$Si and $^{12}$C.
Conclusions are given in Sec. V,
with the outlook of the future extension of our model.
 
\section{Semiclassical distorted wave model description of the $(\pi, K)$
inclusive spectra}
The starting formula for the double differential cross section
in a standard distorted wave model description of the $(\pi ,K)$ hyperon ($Y$)
production inclusive reaction is expressed as
\begin{widetext}
\begin{eqnarray}
\frac{d^2 \sigma}{dW d\Omega} &=& \frac{\omega_i \omega_f}{(2\pi )^2}
 \frac{p_f}{p_i} \int \int d\br d\br ' \sum_{p,h}\;
 \frac{1}{4\omega_i \omega_f}\chi_f^{(-)*}(\br) v_{f,p,i,h}
 \chi_i^{(+)}(\br) \chi_f^{(-)}(\br') 
 v^*_{f,p,i,h} \chi_i^{(+)*}(\br') \nonumber \\
 && \times \phi_p^*(\br) \phi_h (\br) \phi_p(\br') \phi_h^* (\br')
 \delta (W-\epsilon_p + \epsilon_h ) \theta (\epsilon_F -\epsilon_h ),
\end{eqnarray}
\end{widetext}
where $\chi_i^{(+)}$ and $\chi_f^{(-)}$ represent the incident pion and
final kaon wave functions with energies $\omega_i$ and $\omega_f$,
respectively, and $W=\omega_i -\omega_f$ is the energy transfer.
The formula describes the process in which the nucleon in the occupied
single-particle state $h$ is converted to the unobserved outgoing
hyperon ($\Lambda$ or $\Sigma$) state $p$. The elementary amplitude
of the process $\pi + N \rightarrow K + Y$ is denoted
by $v_{f,p,i,h}$, which depends on the energy and momentum of the
particles in the reaction.
In order to treat such dependence, it is
necessary to introduce momentum space integration. In that case,
the explicit calculations involve higher-dimensional integrations.
It is desirable, in practice, to develop a tractable and trustful
approximation method. One procedure that has been frequently used is
the factorization
approximation, in which the elementary process is taken out of the
integration, assuming some averaging wisdom.
As used in Ref. \cite{KEK} for $\pi^- +p\rightarrow K^+ +\Sigma^-$,
the elementary amplitude in the integrand
may be replaced by the averaged differential
cross section over the nucleon momentum distribution $\rho (\bk )$,
\begin{equation}
\overline{\frac{d\sigma(\pi^- p \rightarrow K^+\Sigma^-)}{d\Omega}}
 \equiv \frac{\int \rho (\bk)
\frac{d\sigma}{d\Omega}(\Omega_k)\delta(k-P) dk}
{\int  \rho (\bk)\delta(k-P) dk}
\end{equation}
with $P=k_K +k_Y -k_\pi$, and is taken outside of the integration.
The remaining quantity is the Green function, which is not difficult
to evaluate in the case of a local optical potential. A more sophisticated
Fermi-averaging method was used in Ref. \cite{HH}.
Though such procedure has been widely applied to show various successes,
the justification is far from trivial. Important dynamical effects
might be hidden in the averaging treatment.

\subsection{SCDW method}
In Ref. \cite{MK}, we presented our SCDW approximation method for the
DWIA cross section formula, Eq. (1). There, we introduced a local Fermi
gas approximation for the target nucleus. In this paper, we improve
the description by explicitly treating s.p. wave functions of the target
nucleus.

The semiclassical treatment was first introduced
in the description of the intermediate energy nucleon reactions
on nuclei \cite{SCDW1}. Since the amount of numerical calculations is reduced,
it becomes feasible to include and assess multi-step contributions.
The calculations of $(p,p)$ and $(p,n)$ inclusive spectra
have shown that the SCDW method works well.

The semiclassical approximation employs the following idea for the
propagation of the wave function. Denoting the midpoint and the
relative coordinates of $\br$ and $\br'$ in Eq. (1) by
$\bR= \frac{\br + \br'}{2}$ and $\bs= \br'-\br$, respectively,
we assume that the propagation of the distorted waves, $\chi_i$ and
$\chi_f$, from $\bR$
to $\br$ or $\br'$ is described by a plane wave with the local
classical momentum $\bk(\bR)$ at the position $\bR$.
\begin{eqnarray}
 \chi_i^{(+)}\left(\bR\pm\frac{1}{2}\bs \right) &\simeq &
\mbox{e}^{\pm i\frac{1}{2}\bs \cdot \bk_i (\bR)} \chi_i^{(+)}(\bR), \\
 \chi_f^{(-)}\left(\bR\pm\frac{1}{2}\bs \right)
 &\simeq& \mbox{e}^{\pm i\frac{1}{2}\bs\cdot \bk_f (\bR)} \chi_f^{(-)}(\bR).
\end{eqnarray}
The local momentum $\bk(\bR)$ is defined as follows. The direction
is specified by the quantum mechanical momentum density $\bk_q(\bR)$
calculated by
\begin{equation}
 \bk_q(\bR)= \frac{\Re \{\chi^{(\pm)*}(\bR)(-i)\nabla\chi^{(\pm)}(\bR)\}}
 {|\chi^{(\pm)}(\bR)|^2},
\end{equation}
where $\Re$ represents taking the real part, and the magnitude
is determined by the energy-momentum relation
$\frac{\hbar^2}{2\mu} k^2(\bR) + U_R(\bR) = E$ at $\bR$. Here,
$U_R(\bR)$ is the real part of an optical potential for the
distorted wave function $\chi$ with energy $E$. The relativistic
energy-momentum relation is used for the distorted wave function
described by the Klein-Gordon equation.

The above approximation is expected to work well if the dominant
contributions in the integration over $\br$ and $\br'$ in Eq. (1)
is restricted in the region where $\br$ and $\br'$ are close to
each other. Actually, the density matrix
$\sum_h \phi_h^*(\br')\phi_h(\br)$ brings about this desirable
feature, as is shown in the following heuristic argument.
It is sufficient for the qualitative discussion to assume that
nuclear s.p. wave functions are harmonic oscillator ones.
The summation over the $z$-component of the angular momentum of
each orbit means that we are treating two oscillator functions
coupled to the total angular momentum $L=0$;
\begin{eqnarray}
 & & \sum_{m_h}\; \phi_h^*(\br_1)\phi_h(\br_2) \nonumber \\
& \rightarrow & \sqrt{2\ell_h +1}
 |n_h \ell_h (\br_1),n_h\ell_h (\br_2);L=0\rangle.
\end{eqnarray}
The transformation to the $\bR$ and $\bs$ coordinates are carried out
using the Talmi-Moshinsky brackets.
\begin{eqnarray}
 |n_h \ell_h (\br_1),n_h\ell_h (\br_2);L=0\rangle =\hspace*{3cm}\nonumber\\
  \sum_{N,n,\ell}
  \langle n\ell,N\ell;0|n_h\ell_h,n_h\ell_h;0\rangle 
  |n\ell(\bs),N\ell (\bR);L=0\rangle.
\end{eqnarray}
The reaction processes which we consider take place mostly at the surface
of the target nucleus. When $\bR$ is located in the surface region, the
dominant components in the right hand side of Eq. (7) are those
in which $2N+\ell$ is the largest. This indicates that the dependence
on the relative coordinate $\bs$ is governed by the $0s$
($n=0$ and $\ell =0$) function, which is certainly short-ranged compared
with the size of the target nucleus. Thus we expect that the SCDW
treatment of Eqs. (3,4) in Eq. (1) is meaningful.

Note that since the SCDW approximation should be exact in homogeneous
matter, the SCDW works well inside of the nucleus. The above
reasoning implies that the SCDW approximation is also applicable
to the surface region. This fact is probably connected to the fact
that the local density approximation based on the density matrix
expansion method \cite{NV} works well in nuclear structure
calculations, including the surface region.

\subsection{Wigner transformation}
In the preparatory calculations in Ref. \cite{MK}, we introduced a
Thomas-Fermi approximation for the density
matrix $\sum_h\;\phi_h^*(\br')\phi_h(\br)$ of the target nucleus.
Here, we elaborate the description of the density matrix by using
a Wigner transformation.

The Wigner transformation of the density
matrix of the target nuclear wave function is defined as
\begin{eqnarray}
 \sum_h\;\phi_h^*(\br')\phi_h(\br)
 &=& \sum_h \phi^*_h \left(\bR -\frac{1}{2}\bs \right)
 \phi_h \left(\bR +\frac{1}{2}\bs \right) \nonumber \\
 &=& \int d\bK \;\sum_h \Phi_h (\bR,\bK)\;
e^{i\bK\cdot\bs}.
\end{eqnarray}
$\Phi_h (\bR,\bK)$ is given by the inverse transformation as
\begin{eqnarray}
 \Phi_h (\bR,\bK) &\equiv& \frac{1}{(2\pi)^3} \int d\bs\;e^{-i\bs\cdot\bK}
 \nonumber \\
 && \times \;\phi_h^*(\bR-\frac{1}{2}\bs)\phi_h(\bR+\frac{1}{2}\bs).
\end{eqnarray}
The summation over the $z$-component of the angular momentum is implicit
in these expressions. As is shown in the Appendix, $\Phi_h(\bR,\bK)$ may
be expressed in terms of the Legendre expansion:
\begin{equation}
\Phi_h (\bR,\bK)= \sum_{\ell=\mbox{even}} P_\ell (\cos \widehat{\bK\bR})
\; \Phi_h^\ell (R,K),
\end{equation}
where $\widehat{\bK\bR}$ denotes the angle between two vectors $\bK$
and $\bR$. It is easy to see that the Thomas-Fermi approximation for
the density matrix used in Ref. \cite{MK} amounts to the replacement
\begin{eqnarray}
\sum_h \sum_\ell P_\ell (\cos \widehat{\bK\bR}) \;\Phi_h^\ell (R,K) \nonumber\\
\rightarrow \frac{2}{(2\pi )^3} \theta (k_F(R)-K),
\end{eqnarray}
where $\theta (x)$ is a step function and
$k_F(R)=[3\pi^2 \rho_\tau (R)]^{1/3}$
is a local Fermi momentum determined by the local proton or
neutron density $\rho_\tau (R)$ at $R$.

We describe unobserved hyperons by a local optical potential.
Actually, the hyperon optical potential should be complex,
because there are inelastic processes. The standard way to treat the
completeness of the final states described by a complex Hamiltonian
is to use the Green function method \cite{HLM}. At this stage,
however, we adopt a simplified prescription, employing
a real potential of the standard Woods-Saxon form,
\begin{equation}
 U_Y(r)= U_Y^0/\{1+\exp ((r-r_0)/a)\},
\end{equation}
and we convolute the result of the calculated spectrum with a
Lorentz-type distribution function with the typical width, simulating
the effects of inelastic channels. In that case, the expression of
Eq. (1) is directly used. Then, we introduce the SCDW approximation
as in Eqs. (3,4) also for the hyperon wave functions
$\phi_p(\br)$ and $\phi_p(\br')$.

Using these approximations explained above, the double differential
cross section in the SCDW model becomes
\begin{widetext}
\begin{eqnarray}
\displaystyle
\frac{d^2 \sigma}{dW d\Omega} &=& \frac{\omega_i \omega_f}{(2\pi )^2}
 \frac{p_f}{p_i} \int d\bR \int d\bK \; \sum_p\;
 \frac{1}{4\omega_i \omega_f} |\chi_f^{(-)}(\bR)|^2 |\chi_i^{(+)}(\bR)|^2
 |\phi_p (\bR)|^2  (2\pi )^3 \sum_h \sum_\ell
 P_\ell (\cos \widehat{\bK\bR}) \;\Phi_h^\ell (R,K)\nonumber\\
 && \times |v_{f,p,i,h} (\bK,\bk_i)|^2
  \delta (\bK +\bk_i(\bR) - \bk_f(\bR)- \bk_p(\bR) ) \;
   \delta (W-\epsilon_p +\epsilon_h). \label{eqSCDW}
\end{eqnarray}
\end{widetext}
The summation $\sum_p$ means the sum over the spin and the momentum of
the outgoing unobserved hyperon: $\frac{1}{(2\pi )^3}\sum_{\mbox{spin}}
 \int \:d\bp$. If we use the energy, instead of the momentum,
to specify scattering states, the momentum integration $\int \:d\bp$
is written as follows:
\begin{equation}
 \displaystyle \int
 \;d\Omega_p \int \frac{1}{2}
\left(\frac{2m_Y}{\hbar^2} \right)^{3/2} \sqrt{\epsilon_p }\; d\epsilon_p.
\end{equation}

The above final expression (\ref{eqSCDW}) admits of the simple interpretation
that the reaction in which $\pi +N$ yields $K+Y$ takes place at the position
$\bR$ and satisfies conservation of local semiclassical momentum:
\begin{equation}
 \bK +\bk_i(\bR) = \bk_f(\bR)+ \bk_p(\bR)
\end{equation}
These momenta, $\bk_i(\bR)$, $\bk_f(\bR)$ and $\bk_p(\bR)$, are calculated
with Eq. (5), using $\pi$, $K$ and $Y$ distorted wave functions in
an optical model description. It is seen that the dimension of the
integration does not change from Eq. (1) to Eq. (\ref{eqSCDW}).
However, we can now treat the momentum dependence of the transition
amplitude $v_{f,p,i,h}$ explicitly.

In the present formulation, the correction for the lack of the translational
invariance \cite{ES} in the target nuclear wave function is not included.
In describing electron scattering on a nucleus composed of $A$ nucleons,
the center-of-mass effects \cite{TB} in the nuclear shell model have been
taken care of by a multiplicative factor $F^{1/2}=\exp (q^2/(4A\alpha))$,
where $q$ is the momentum transfer and $\alpha$ is the oscillator constant.
For $A=12$ and $\alpha=0.4$ fm$^{-2}$, the factor $|F^{1/2}|^2$ amounts to
1.5 when the momentum transfer is 2 fm$^{-1}$. Thus the calculation without
the center-of-mass correction tends to underestimate the cross section.
The appropriate treatment of the center-of-mass effects in our SCDW method
deserves to be studied in the future.

\subsection{Elementary amplitude}
The on-shell amplitude $v_{f,p,i,h}$ of the elementary process is
related to the differential cross section by
\begin{equation}
 \frac{d\sigma}{d\Omega} = \frac{1}{(4\pi)^2} \frac{E_N E_Y}{s}
 \frac{k_K}{k_\pi} |v|^2,
\end{equation}
where $s$ is the invariant mass squared. In Eq. (\ref{eqSCDW}),
we are able to account for the angular dependence of
the $\pi+ N \rightarrow K +Y$ elementary process at each
position $\bR$. To carry out an actual calculation in Eq. (\ref{eqSCDW}),
we need some model description for $v$. However, at present, there is
no reliable model for the relevant process, including off-shell regions.
We use a simple phenomenological parametrization based on Eq. (16),
by defining the invariant mass squared using the momenta $\bk_i(\bR)$
and $\bK$.
The following functional form is used to simulate empirical angular
distributions of the  $\pi+ N \rightarrow K +Y$ reactions.
\begin{equation}
 \frac{d\sigma}{d\Omega} = \frac{\sigma (\sqrt{s})}{4\pi}
 \{ 1+ \sum_\ell a_\ell(\sqrt{s}) P_\ell (\cos \theta )\}
\end{equation}
Values of $\sigma (\sqrt{s})$ and $a_\ell(\sqrt{s})$ used in
the present calculations will be
given in Sec. IV.

\subsection{Wave functions of the target nucleus and hyperons}
In this paper, we are concerned with the reactions on the $N=Z$
targets, $^{28}$Si and $^{12}$C. Single-particle wave functions and
the energies for these nuclei $\epsilon_h$ are prepared by the
density-dependent Hartree-Fock description of Campi-Sprung \cite{CS}.

As stated in {\bf B} of this section, hyperons are described by an
energy-independent local potential of the Woods-Saxon form.
We use the standard geometry parameters,
$r_0=1.25 \times (A-1)^{1/3}$ fm and $a=0.65$ fm. The Coulomb
potential regularized in a nucleus is incorporated in the case
of the $\Sigma^-$. It is noted that if we use a different parameter set
such as $r_0=1.20 \times (A-1)^{1/3}$ fm and $a=0.6$ fm, we do
not see appreciable changes in calculated spectra for $^{12}$C and
$^{28}$Si.

It may be argued that the hyperon s.p. potential is not expressed
by a single Woods-Saxon shape. The $\Sigma$ potential may be
repulsive in the higher density region, but change its sign at
the surface as has been suggested by the analysis \cite{BFG} of
$\Sigma^-$ atomic data and also by the nuclear matter calculations
\cite{KOH} with the SU(6) quark model interaction. At present,
however, it is premature to discuss the detailed shape of the $\Sigma$
potential from the available inclusive spectra. We assume,
from the beginning, the standard Woods-Saxon form and ask what
strength $U_\Sigma^0$ is favored by the experimental data.
More experimental data is needed to quantitatively
discuss more elaborate shape parameters as well as the possible
energy-dependence of the strength.

The nuclear matter calculations \cite{KOH} with the SU(6) quark model
interaction suggest that the imaginary strength of the $\Lambda$ s.p.
potential hardly depends on the density, but that of the $\Sigma$ s.p.
potential decreases when the nuclear matter density
becomes lower. Regarding that the hyperon formation processes take
places at the surface region, we set the energy dependence of the
half width in MeV for smearing the calculated spectra as follows,
simulating the calculated results at the half of the normal density
with the quark model potential FSS:
\begin{eqnarray}
 \frac{\Gamma_\Lambda (E)}{2}&=& \left\{ \begin{array}{l}
 1+12(E/40)^2-8(E/40)^3, \hspace{0.5em}E\le 40 \mbox{ MeV} \\
 \\
 5, \hspace{1em}E> 40 \mbox{ MeV},
 \end{array} \right. \\
 \frac{\Gamma_\Sigma (E)}{2} &=& 10+\frac{10}{1+(20/E)^2}, \hspace{0.5em}
 \mbox{$E$ in MeV}.
\end{eqnarray}

\section{Optical potentials for pions and kaons}

\begin{figure}[t]
\epsfxsize=8.0cm
\epsfbox{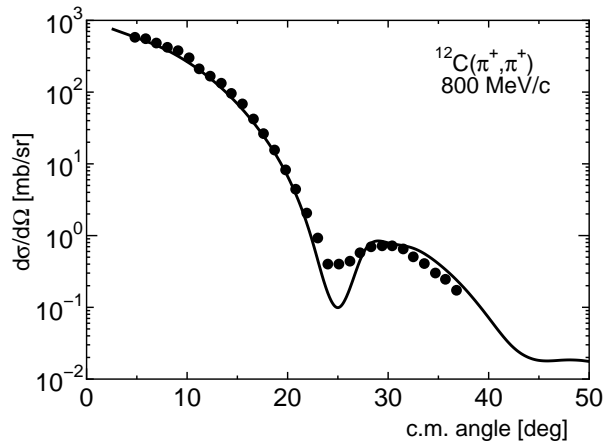}
\caption{Differential cross section of $\pi^+$ elastic scattering
on $^{12}$C at $p_\pi =800$ MeV/c. Calculated values in the optical
potential model with $-i b_0=1.0$ fm$^3$ is compared with the
experimental data \cite{CPI}.
}
\end{figure}

The incident pions and detected kaons are described by the standard
Klein-Gordon equation with some optical potential model.
Following the usual procedure to construct a $\pi$-nucleus optical
potential from $\pi N$ elementary amplitudes, the optical potential for
the pion is given by
\begin{equation}
 V_{\pi}(r)= - \frac{k^2}{2E_\pi} b_0 \rho (r)
\end{equation}
where $\rho (r)$ is the one-body nuclear density distribution and
the parameter $b_0$ is related to the sum of isospin averaged $\pi N$
partial wave amplitudes. In practice, a pure imaginary choice of
$b_0=i\frac{1}{k}\langle \sigma_{tot}\rangle$ is found to work well.
As an example, we show, in Fig. 1, the pion elastic scattering
differential cross sections on $^{12}$C at 800 MeV/c.
We simply expect at the present stage that the same prescription
is applicable to the incident momentum of 1.2 GeV/c. Actually,
$-i b_0=0.58$ fm$^3$ at 1.2 GeV/c and $-i b_0=0.70$ fm$^3$ at 1.05 GeV/c.

\begin{figure}[t]
\epsfxsize=6.6cm
\epsfbox{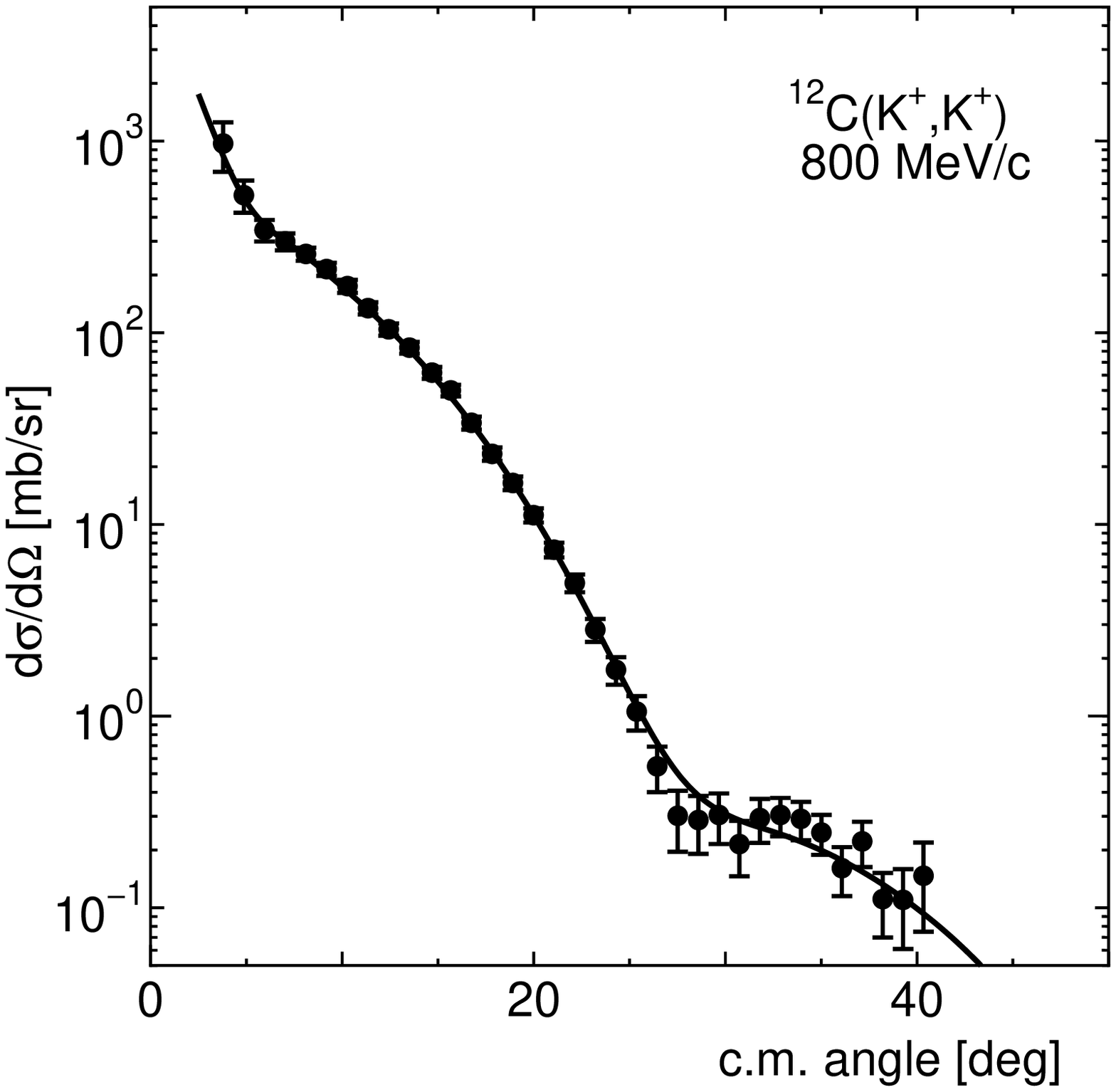}
\caption{Differential cross section of $K^+$ elastic scattering
on $^{12}$C at $p_K=800$ MeV/c. The optical model with
the parameter $b_0$ given in Eq. (21) gives the
result shown by a solid curve, compared with the experimental
data \cite{MAR}.
}
\bigskip
\epsfxsize=6.6cm
\epsfbox{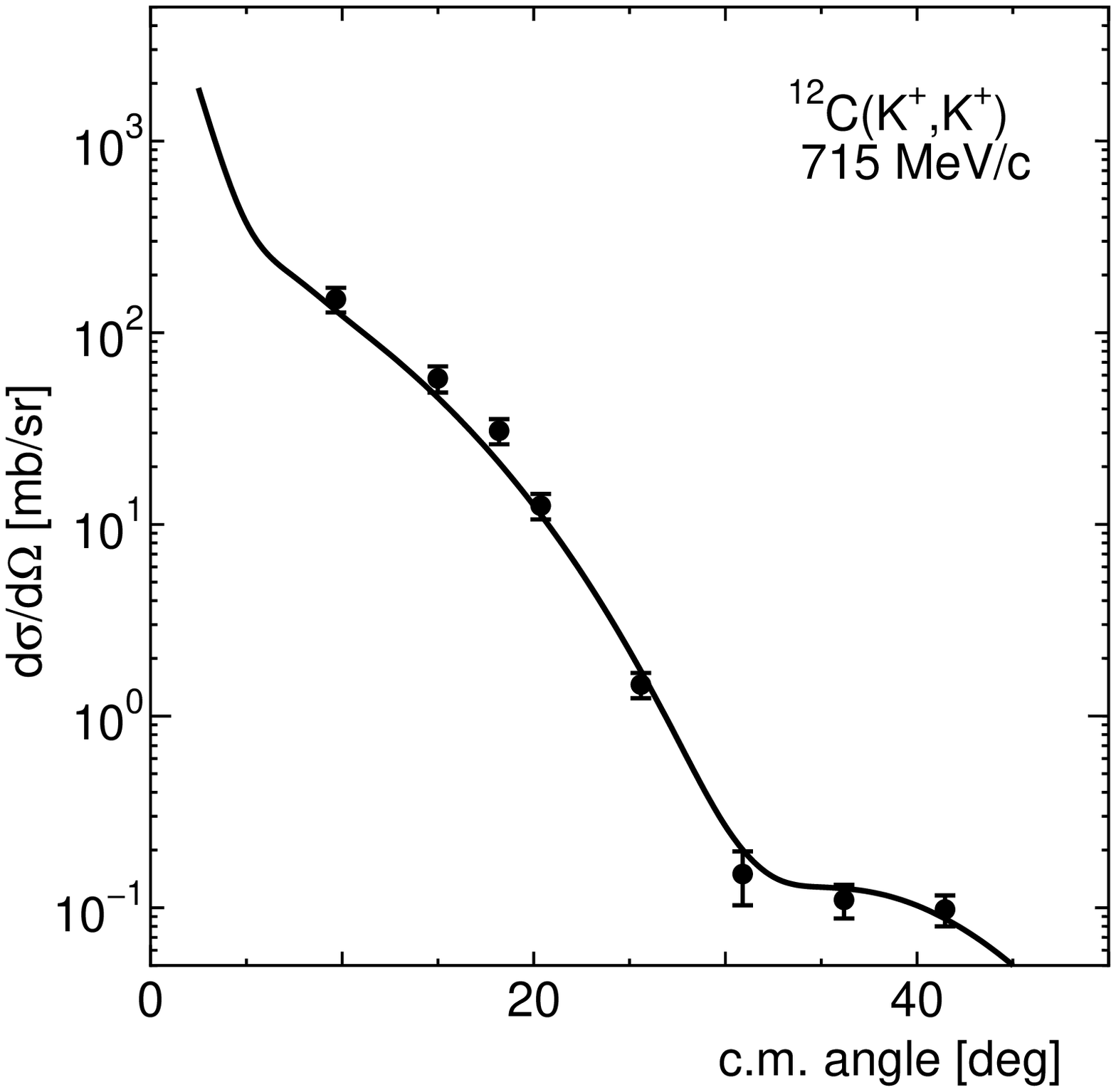}
\caption{Differential cross section of $K^+$ elastic scattering
on $^{12}$C at $p_K =715$ MeV/c. The optical model with
the parameter $b_0$ given in Eq. (21) gives the
result shown by a solid curve, compared with the experimental
data \cite{MIC}.
}
\end{figure}

It has been known \cite{SKG} that the kaon scattering data is not well
reproduced by simply folding elementary amplitudes. In the present
calculation, we phenomenologically search an optimal parameter $b_0$
in the form of Eq. (20), using the available experimental
data \cite{MAR,MIC} on C at 800 MeV/c and 715 MeV/c. For the K$^+$,
we find that the following momentum dependence is adequate;
\begin{equation}
 b_0= -6.2 \times 10^{-4} p_K +i \frac{90.0}{p_K},
\end{equation}
where $b_0$ in fm$^3$ and $p_K$ in MeV/c.
As shown in Figs. 2 and 3, the calculated differential cross sections
account well for the experimental data. The outgoing $K^+$ momenta
relevant to the present inclusive spectra are in this energy region.

\begin{table}
\caption{Legendre coefficients $a_\ell (\sqrt{s})$ in Eq. (17) of
the differential cross sections of
the $\pi^- + p \rightarrow K^0 + \Lambda$ reaction,
determined using experimental data \cite{BAK, SAX}.}
\vspace*{2mm}

\begin{tabular}{|r|rcl|} \hline
  & $a_\ell$\hspace*{2em} & & \hspace*{1em}range of $\sqrt{s}$ (GeV)\\ \hline
  $\ell =1$ & $12.6\sqrt{s}-20.362$ & & $\sqrt{s} \leq 1.69$ \\
            & $0.932$ &  & $1.69 < \sqrt{s} \leq 1.925$ \\
            & $-4.8\sqrt{s}+10.172$ &  & $1.925<\sqrt{s} \leq 2.015$ \\
            & $2.164\sqrt{s}-3.86$ &  & $2.015<\sqrt{s} \leq 2.3$ \\ \hline
  $\ell =2$ & $0$ & & $\sqrt{s} \leq 1.677$ \\
            & $11.82\sqrt{s}-19.821$ &  & $1.677 < \sqrt{s} \leq 1.7$ \\
            & $0.273$ &  & $1.7<\sqrt{s} \leq 1.8$ \\
            & $4.71\sqrt{s}-8.205$ &  & $1.8<\sqrt{s} \leq 2.4$ \\ \hline
  $\ell =3$ & $0$ & & $\sqrt{s} \leq 1.8$ \\
            & $-5.68\sqrt{s}+10.224$ &  & $1.8 < \sqrt{s} \leq 1.85$ \\
            & $-0.284$ &  & $1.85<\sqrt{s} \leq 1.98$ \\
            & $-8.32\sqrt{s}+16.19$ &  & $1.98<\sqrt{s} \leq 2.03$ \\
            & $6.13\sqrt{s}-13.144$ &  & $2.03<\sqrt{s} \leq 2.4$ \\ \hline
  $\ell =4$ & $0$ & & $\sqrt{s} \leq 1.86$ \\
            & $7.5\sqrt{s}-13.95$ &  & $1.86 < \sqrt{s} \leq 1.96$ \\
            & $0.75$ &  & $1.96<\sqrt{s} \leq 2.08$ \\
            & $5.5\sqrt{s}-10.69$ &  & $2.08<\sqrt{s} \leq 2.4$ \\ \hline
\end{tabular}
\end{table}

\section{Results}
\subsection{$\Lambda$ formation}
We first apply our SCDW model to the $(\pi^+,K^+)$ $\Lambda$ formation
inclusive spectra obtained with $^{28}$Si and $^{12}$C targets
measured at KEK \cite{SAHA,SAHA2}. Since the $\Lambda$ s.p. potential
has been established as $V_\Lambda^0 \sim -30$ MeV from various
$\Lambda$ hypernuclear data \cite{LAMP}, the calculation serves as
quantitative assessment of the SCDW model description.
The strength and angular dependence of the elementary process are
parameterized as Eq. (17) according to the available experimental
data of $\pi^- + p \rightarrow K^0 + \Lambda$ \cite{BAK, SAX}.
The energy dependence of the total cross section is fitted as
\begin{equation}
\sigma(\sqrt{s})=\frac{472 (\sqrt{s}-1.60935)^{1.2}}
{[26(\sqrt{s}-1.60935)]^3+20}\hspace{0.5em}\mbox{[mb]}.
\end{equation}
Coefficients $a_\ell (\sqrt{s})$ for the angular dependence
are tabulated in Table I.

Calculated spectra with three choices of the strength of the $\Lambda$
s.p. potential, $V_\Lambda^0 =-40$, $-30$, $-10$ MeV, are compared
to see the potential dependence. Figure 4 shows the results for $^{28}$Si
with $p_\pi = 1.2$ GeV/c, while Figs. 5 and 6 for $^{12}$C with
$p_\pi = 1.2$ and 1.05 GeV/c, respectively.

\begin{figure}[t]
\epsfxsize=8.0cm
\epsfbox{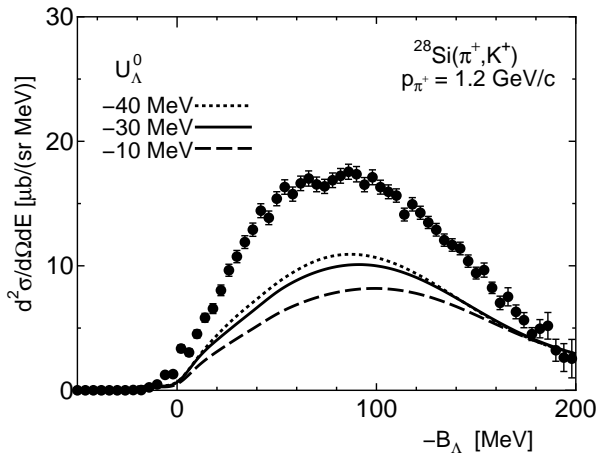}
\caption{$(\pi^+,K^+)$ $\Lambda$ formation inclusive spectra with a $^{28}$Si
target at $\theta_K =6^\circ \mp 2^\circ$ for pions with $p_\pi =1.2$ GeV/$c$.
These results were obtained with three choices of $U_\Lambda^0$ in a
Woods-Saxon potential form with the geometry parameters of $r_0=1.25\times
(A-1)^{1/3}$ fm and $a=0.65$ fm. The KEK data \cite{SAHA2} are also displayed.
}
\end{figure}

The $\Lambda$ hyperon can be bound in a nucleus. In $^{11}$C, $0s$
and $0p$ $\Lambda$ bound states appear at energies of $-12.4$ MeV
and $-1.8$ MeV, respectively, for the case of $V_\Lambda^0 =-30$ MeV
with $r_0=1.25\times 11^{1/3}$ fm and $a=0.65$ fm, neglecting the small
spin-orbit component. The $\Lambda$ bound state wave function is treated
in the same manner as for the target nuclear wave function, Eq. (9).
The scattering wave function $\phi_p$ in Eq. (13) is replaced by
the Wigner transformation of the $\Lambda$ bound state wave function.
The transition strength from the $0p_{3/2}$ nucleon hole state appears
as two peaks below $-B_\Lambda =0$, for which the experimental
resolution of 2 MeV in FWHM is applied. In this presentation, we use
s.p. neutron energies in a simple $0p_{3/2}$-closed HF
description for the target nucleus $^{12}$C. Thus the peak position
does not precisely agree with that of the experimental spectrum.
Since the $0s_{1/2}$ nucleon hole state has a large width, we broaden
the calculated strength by the half width of 8 MeV.

These figures indicate that the energy dependence of the inclusive
spectrum is well reproduced by the standard choice of
the $\Lambda$ s.p. potential, $V_0=-30$ MeV.
The absolute values are short by about 35 \%. As is noted in the
end of Sec. II B, our calculation tends to underestimate the cross section
for lack of the translational invariance in the target wave function.
We also expect various other effects for the underestimation. Besides
possible ambiguities in the SCDW treatment and uncertainties in
the elementary strengths as well as the optical potential parameters,
there should be room for contributions from two-step processes and more.
Since the incident pion absorption is rather large, some
of the flux lost may emerge again into the $\Lambda$ production
channel. Bearing in mind these points and in addition the possible
modification of the elementary process in nuclear medium, which is
clearly premature to discuss at the present stage of the analysis,
our SCDW model is considered to provide a meaningful
description for the $(\pi,K)$ inclusive spectra. The explicit estimation
of the center of mass correction and the two-step contributions is
needed to establish the quantitative reliability of the SCDW method.

\begin{figure}
\epsfxsize=8.0cm
\epsfbox{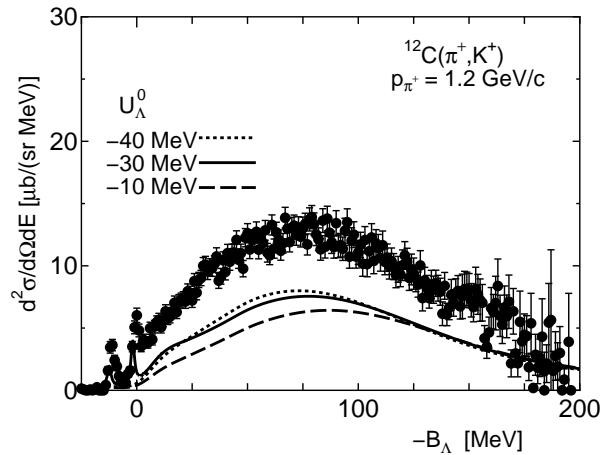}
\caption{$(\pi^+,K^+)$ $\Lambda$ formation inclusive spectra with a $^{12}$C
target at $\theta_K =6^\circ \mp 2^\circ$ for pions with $p_\pi =1.2$ GeV/$c$.
These results were obtained with three choices of $U_\Lambda^0$ in a
Woods-Saxon potential form with the geometry parameters of $r_0=1.25\times
(A-1)^{1/3}$ fm and $a=0.65$ fm. The KEK data \cite{SAHA,SAHA2} are also
displayed.
}
\end{figure}

\begin{figure}
\epsfxsize=7.6cm
\epsfbox{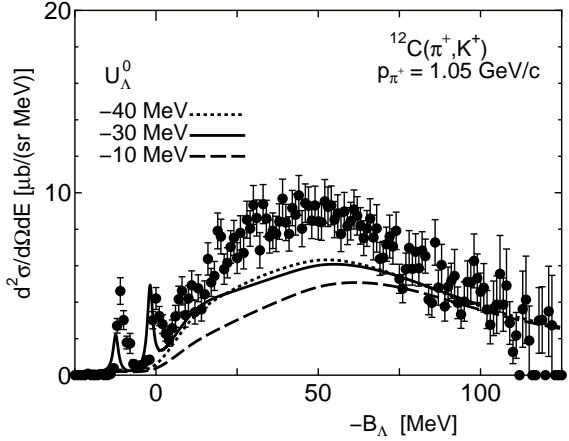}
\caption{$(\pi^+,K^+)$ $\Lambda$ formation inclusive spectra with a $^{12}$C
target at $\theta_K =6^\circ \mp 2^\circ$ for pions with $p_\pi =1.05$ GeV/$c$.
These results were obtained with three choices of $U_\Lambda^0$ in
a Woods-Saxon potential form with the geometry parameters of $r_0=1.25\times
(A-1)^{1/3}$ fm and $a=0.65$ fm. The KEK data \cite{SAHA2} are also displayed.
}
\end{figure}

\subsection{$\Sigma$ formation}
In Ref. \cite{MK}, we assumed an isotropic angular dependence for
the $\pi^- + p\rightarrow K^++\Sigma^-$ elementary process in calculating
$(\pi^-,K^+)$ $\Sigma^-$ formation inclusive spectra. The energy
dependence of $\sigma(\sqrt{s})$ was taken from the parameterization
by Tsushima {\it et al.} \cite{THF}, which was renormalized by a factor
of 0.82. In this paper, we take into account the angular distribution
using the Legendre polynomial coefficients
reported by Good and Kofler \cite{GK} on the basis of available data
\cite{GK,GOU,DAHL,DOY}. Up to $\sqrt{s}\sim 2.1$ GeV,
$p_\pi \sim 1900$ MeV/c, we can set $a_\ell =0$ with $\ell \ge 3$.
We try to simulate the energy dependence of $a_1(\sqrt{s})$ and
$a_2(\sqrt{s})$ by several lines as given in Table II, which are depicted
in Fig.7 with the experimental data.
The energy dependence of the total cross section $\sigma(\sqrt{s})$
is parameterized as follows:
\begin{eqnarray}
\sigma(\sqrt{s})&=& \frac{0.02055 (\sqrt{s}-1.691)^{0.9603}}
{(\sqrt{s}-1.682)^2+0.003131} \nonumber \\
 & +& \frac{0.003023 (\sqrt{s}-1.691)^{0.1394}}{(\sqrt{s}-1.894)^2+0.01548} 
 \hspace{0.5em}\mbox{[mb]},
\end{eqnarray}
which is displayed by the solid curve in Fig. 8.
At $\sqrt{s}=1.79$ GeV, corresponding to $p_\pi = 1.2$ GeV/c,
the absolute magnitude of the differential cross section at forward angles
in the laboratory frame is about 130 $\mu$b, which corresponds well to
that measured at KEK \cite{KEK}.

\begin{table}
\begin{center}
\caption{Legendre coefficients $a_\ell (\sqrt{s})$ in Eq. (17) of
the differential cross sections of
the $\pi^-+ p \rightarrow K^+ + \Sigma^-$ reaction,
determined using experimental data \cite{GK,GOU,DAHL,DOY}.}
\vspace*{2mm}

\begin{tabular}{|r|rcl|} \hline
  & $a_\ell$\hspace*{2em} & & \hspace*{1em}range of $\sqrt{s}$ (GeV) \\ \hline
  $\ell =1$ & $0.0$ & & $\sqrt{s} \leq 1.72$ \\
            & $-7.134(\sqrt{s}-1.72)$ &  & $1.72 < \sqrt{s} \leq 1.93$ \\
            & $-1.5$ &  & $1.93 <\sqrt{s} \leq 1.97$ \\
            & $6.25\sqrt{s}-13.81$ &  & $1.97<\sqrt{s} \leq 2.05$ \\
            & $-1.0$ &  & $2.05 <\sqrt{s} \leq 2.15$ \\
            & $-4.06\sqrt{s}+7.729$ &  & $2.15 <\sqrt{s} \leq 2.31$ \\
            & $-1.65$ &  & $2.31 <\sqrt{s} $ \\ \hline
  $\ell =2$ & $0.55$ & & $\sqrt{s} \leq 1.81$ \\
            & $8.0\sqrt{s}-13.93$ &  & $1.81 < \sqrt{s} \leq 1.89$ \\
            & $-7.44\sqrt{s}+15.25$ &  & $1.89<\sqrt{s} \leq 2.05$ \\
            & $0.0$ &  & $2.05<\sqrt{s} \leq 2.15$ \\
            & $10.0\sqrt{s}-21.5$ &  & $2.15<\sqrt{s} \leq 2.31$ \\
            & $1.6$ &  & $2.31<\sqrt{s}$ \\
             \hline
\end{tabular}

\end{center}
\end{table}

\begin{figure}
\epsfxsize=7.6cm
\epsfbox{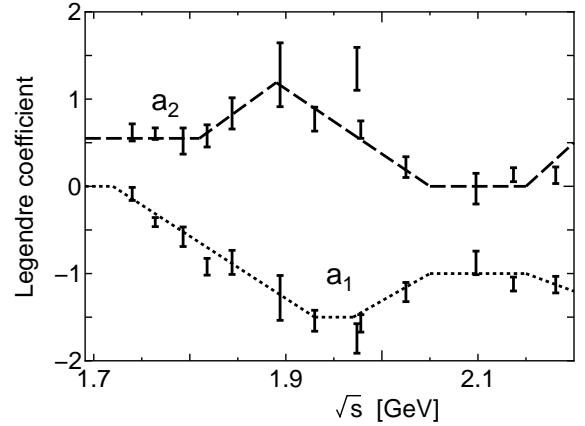}
\caption{$\sqrt{s}$-dependence of Legendre coefficients of angular
distributions of the $\pi^- + p\rightarrow K^++\Sigma^-$ reaction.
Empirical data points are taken from Refs. \cite{GK,GOU,DAHL,DOY}.
The dotted and dashed lines show the fit for the
calculation of $(\pi^-,K^+)$ inclusive spectra.
}
\end{figure}

\begin{figure}
\epsfxsize=7.6cm
\epsfbox{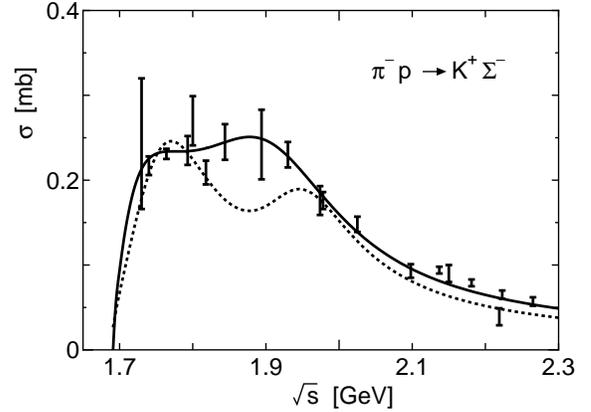}
\caption{Total $\pi^- + p\rightarrow K^++\Sigma^-$ cross sections as a function
of the c.m. energy. Data points are taken from Refs. \cite{GK,GOU,DAHL,DOY}.
The solid curve displays the parameterization of Eq. (23) in the present
calculation. For comparison, the energy dependence used in Ref. \cite{MK} with
an isotropic angular dependence is shown by the dotted curve. 
}
\end{figure}

\begin{figure}[t]
\epsfxsize=8.0cm
\epsfbox{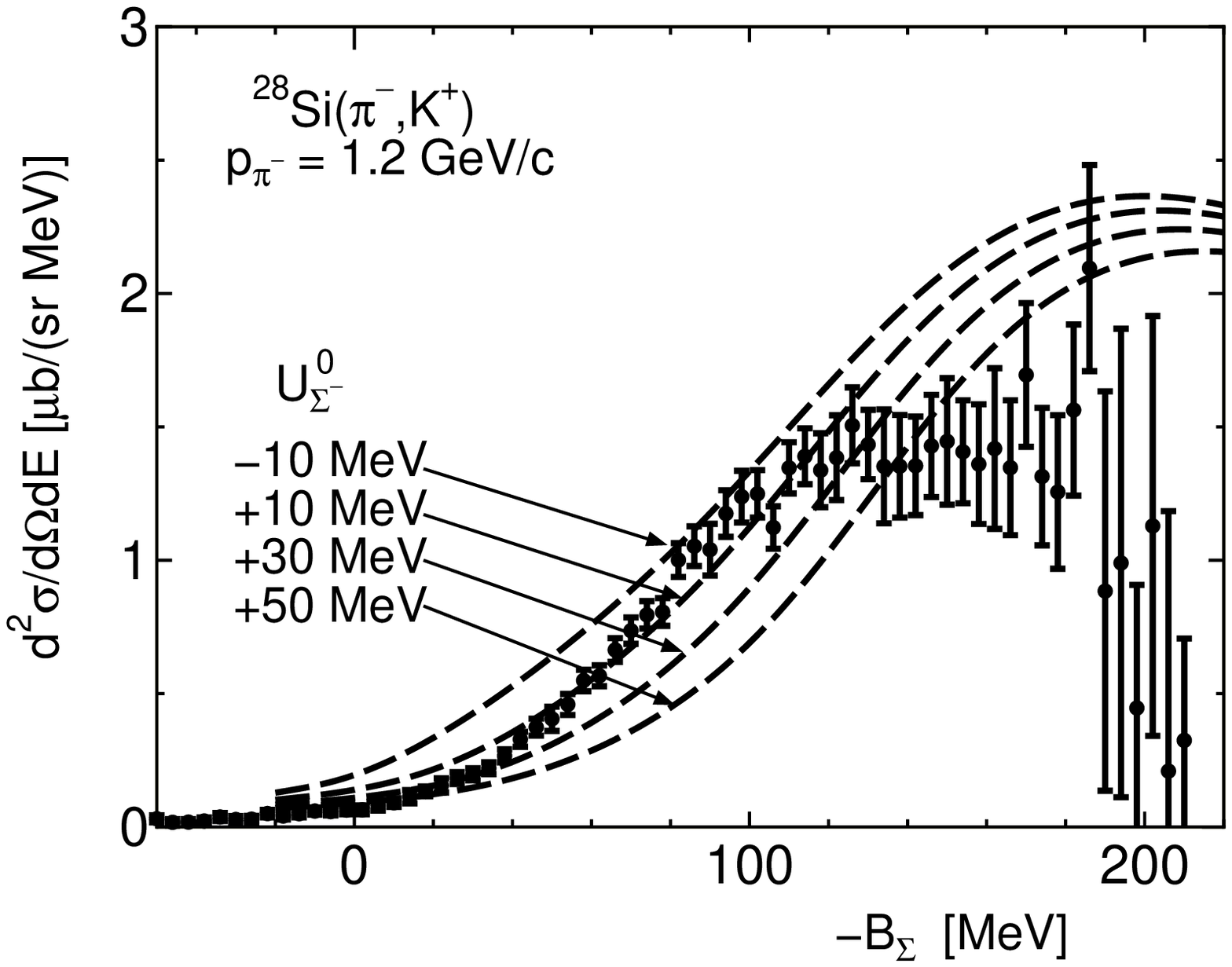}
\caption{$(\pi^-,K^+)$ $\Sigma$ formation inclusive spectra with a $^{28}$Si
target at $\theta_K =6^\circ \mp 2^\circ$ for pions with $p_\pi =1.2$ GeV/$c$.
These results were obtained with four choices of the repulsive strength 
$U_\Sigma^0 = -10, 10, 30, 50$ in a Woods-Saxon potential form with the
geometry parameters of $r_0=1.25\times (A-1)^{1/3}$ fm and a=0.65 fm.
Experimental data points are taken from Refs. \cite{KEK,SAHA}.
}
\end{figure}

\begin{figure}
\epsfxsize=8.0cm
\epsfbox{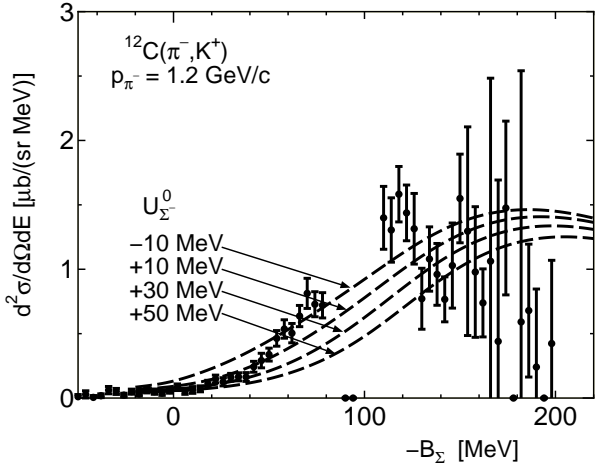}
\caption{$(\pi^-,K^+)$ $\Sigma$ formation inclusive spectra with a $^{12}$C
target at $\theta_K =6^\circ \mp 2^\circ$ for pions with $p_\pi =1.2$ GeV/$c$.
These results were obtained with four choices of the repulsive strength 
$U_\Sigma^0 = -10, 10, 30, 50$ in a Woods-Saxon potential form with the
geometry parameters of $r_0=1.25\times (A-1)^{1/3}$ fm and a=0.65 fm.
Experimental data points are taken from Refs. \cite{SAHA}.
}
\end{figure}

Figures 9 and 10 compare calculated $(\pi^-, K^+)$
inclusive spectra with the KEK experimental data \cite{KEK,SAHA}.
Several curves in these figures correspond to the assumed $\Sigma$
potential with $U_\Sigma^0=-10$, $10$, $30$ and $50$ MeV. In DWIA
analyses in Refs. \cite{KEK,SAHA} and also in Ref. \cite{HH}, an overall
reduction factor is introduced to discuss the correspondence with
the experimental data. However, we do not
multiply any renormalization factor. It is seen that absolute values
are satisfactorily reproduced by a repulsive strength of $10\sim 30$ MeV.
Since we may expect additional contributions
from multi-step processes, the actual repulsive strength may be larger
than 30 MeV. This result agrees with that in Ref. \cite{MK}, where
the $\Sigma$ potential was concluded to be repulsive of the order
of $30\sim50$ MeV, using the SCDW model with the Thomas-Fermi
approximation for the target nucleus, $^{28}$Si. The assumption of
the isotropic angular distribution of the elementary process used
in Ref. \cite{MK} tends to overestimate the elementary cross section
at forward angles. Thus, the repulsive strength needed to reproduce
the experimental data becomes smaller.

\begin{table}
\begin{center}
\caption{Percentage contribution from each c.m. energy ($\sqrt{s}$)
region of the $\pi^-+ p \rightarrow K^+ + \Sigma^-$ elementary process
at $-B_\Sigma$ = 50, 110 and 170 MeV in $(\pi^-,K^+)$ $\Sigma$ formation
inclusive spectra on $^{28}$Si with $U_\Sigma^0=10$ and 30 MeV, respectively.
}
\vspace*{2mm}

\begin{tabular}{|c|rrr|} \hline
     \multicolumn{4}{|c|}{case of $U_\Sigma^0=10$ MeV} \\ \hline 
 &  \multicolumn{3}{c|}{$-B_\Sigma$ (MeV)} \\
  \hspace*{1em}range of $\sqrt{s}$ (GeV) &   50  & 110  & 170  \\ \hline
          $\sqrt{s}\leq 1.75$   &    0.0 \% &    0.0 \%  &   3.7 \%  \\
 1.75 $< \sqrt{s}\leq 1.80$   &    0.0 \%  &  25.7 \%  &  68.2 \% \\
 1.80 $< \sqrt{s}\leq 1.85$   &  21.8 \%  &  62.7 \%  &  27.3 \% \\
 1.85 $< \sqrt{s}\leq 1.90$   &  68.4 \%  &  11.2 \%  &   0.8 \% \\
 1.90 $< \sqrt{s}\leq 2.00$   &   9.7 \%  &   0.4 \%   &  0.0 \%  \\
 2.00 $< \sqrt{s}$           &   0.1 \%   &   0.0 \%   &  0.0 \%  \\ \hline \hline
      \multicolumn{4}{|c|}{case of $U_\Sigma^0=30$ MeV} \\ \hline 
     &  \multicolumn{3}{c|}{$-B_\Sigma$ (MeV)} \\
  \hspace*{1em}range of $\sqrt{s}$ (GeV) &   50  & 110  & 170  \\ \hline
          $\sqrt{s}\leq 1.75$   &    0.0 \% &    0.0 \%  &   1.8 \%  \\
 1.75 $< \sqrt{s}\leq 1.80$   &    0.0 \%  &  18.1 \%  &  48.3 \% \\
 1.80 $< \sqrt{s}\leq 1.85$   &  12.2 \%  &  55.2 \%  &  46.2 \% \\
 1.85 $< \sqrt{s}\leq 1.90$   &  66.6 \%  &  25.0 \%  &   3.6 \% \\
 1.90 $< \sqrt{s}\leq 2.00$   &  20.9 \%  &   1.7 \%   &  0.1 \%  \\
 2.00 $< \sqrt{s}$           &   0.3 \%   &   0.0 \%   &  0.0 \%  \\ \hline
\end{tabular}

\end{center}
\end{table}

At present, the agreement of the calculated shape of the spectrum with the
experimental data is not excellent. This may be related to the uncertainties
of the input cross section, besides multi-step contributions. As is seen in
Figs. 7 and 8, error bars of the elementary cross section are rather large.
It is also to be born in mind that the $\Sigma$-nucleus potential may be
energy-dependent. On the experimental side, the data is uncertain at the
higher excitation energy region due to the spectrometer
acceptance \cite{SAHA}.

In order to learn the role of the nucleon Fermi motion,
it is instructive to show which energy region of the elementary process
dominantly contributes to the formation strength at each $\Sigma$ separation
energy, $-B_\Sigma$. Table III tabulates percentage contributions from six
different regions of $\sqrt{s}$ at  $-B_\Sigma$ = 50, 110 and
170 MeV for the cases of $U_\Sigma^0=10$ and 30 MeV, respectively.
At lower $\Sigma$ excitation energies, the reaction mainly occurs with
a nucleon moving toward the incident pion. On the other hand, at higher
$\Sigma$ energies, the dominant contribution comes from the elementary
process at lower c.m. energies.

It was remarked in Ref. \cite{KEK} that the peak position of the
$(\pi^-,K^+)$ $\Sigma^-$ formation inclusive spectrum at an energy as
high as 120 MeV is difficult to reproduce if the repulsion of the
$\Sigma$-nucleus potential is not so strong as about 100 MeV.
Our analysis suggests, however, that it is not necessary for the
$\Sigma$ s.p. potential to be such repulsive.
The reason that the result obtained in our SCDW model differs from
that of Ref. \cite{KEK} might be related to the fact that we did not use
the factorization approximation represented by the average cross
section, Eq. (2). The descriptions of incident pion and outgoing kaon
distorted waves are also different. It will be worthwhile to detect
the source of the different results.

It is necessary to discuss the correspondence of the $\Sigma$ potential
strength obtained on the basis of the present $(\pi^-,K^+)$ inclusive
spectra with the predictions of varying theoretical models for the $\Sigma N$
interaction. Most models \cite{NIJDF,NIJNS} of the Nijmegen group give
an attractive $\Sigma$ s.p. potential, as is seen in various $G$-matrix
calculations \cite{YB1,YB2,SCHU,VIDA} in nuclear matter.
Only the model F is acceptable, which was concluded
also by D\c{a}browski \cite{DAB} in his plane wave impulse approximation
analysis of the BNL data
\cite{BNL} of the $(K^-,\pi^+)$ spectrum on $^9$Be at $p_K= 600$ MeV/c.
The recent SU(6) quark model \cite{FU96a,FU96b,FU01} by Kyoto-Niigata
group definitely predicts a repulsive $\Sigma$ s.p. potential. The $G$-matrix
calculations \cite{KOH} in symmetric nuclear matter showed that the early
version, FSS \cite{FU96a,FU96b}, gives
$U_\Sigma^0 \sim +20$ MeV and the latest version, fss2 \cite{FU01},
$U_\Sigma^0 \sim +8$ MeV. This repulsive $\Sigma$ s.p. potential
originates from a strongly repulsive character in the isospin
$T=\frac{3}{2}$ channel due to the quark anti-symmetrization effect.
If the strength of more than 30 MeV is confirmed in future, these
theoretical models will need fine tuning.

It is also important to pay attention to the density dependence
of the $\Sigma$ s.p. potential. As the $\Sigma^-$ atomic data tells
\cite{BFG}, the $\Sigma$ potential is attractive at the very surface region
of a nucleus. This feature is also seen in the $G$-matrix calculation
with the quark model potential \cite{KOH}. 
The calculation in this paper assumes a single Woods-Saxon form for
the $\Sigma$ potential. A question whether the sign change of the
$\Sigma^-$ potential can be detected or not at the outside region in the
description of the $(\pi^-,K^+)$ reaction is to be studied in future. 
The energy dependence of the $\Sigma$ potential is another issue
to be addressed. The $G$-matrix calculation in Ref. \cite{KOH}
indicates that the repulsive strength is not energy-independent.

Because of the strong repulsive contribution from the isospin
$T=\frac{3}{2}$ channel, it is hypothesized that the $\Sigma^-$-nucleus
potential becomes more repulsive in the case
of a neutron excess. In this respect, analyses of the $(\pi^-,K^+)$ data
with heavier nucleus targets will be interesting for the purpose
of investigating whether such quantitative isospin dependence
actually exists.

In the present calculations, there are various treatments to be improved.
The smearing caused by the Lorentz-type convolution should
be treated by the precise way of the Green's function method, though
much calculational efforts have to be devoted. A quantitative estimation
of the contribution from multi-step processes is needed, which could fill
the difference of the experimental data and the calculational results as
is seen in the $\Lambda$ formation spectra.
A model description of the elementary process
is to be upgraded, though new experimental data is required
to do it. After improving these points and the proper account of the
CM motion of the target wave function, the SCDW framework serves as
a quantitatively reliable model to study the possible modification of
the elementary amplitudes in nuclear medium.
The direction of the change of cross sections, increase or decrease,
depends on how the amplitudes are altered through the underlying
dynamical processes.

\section{Conclusions}
We have developed a semiclassical distorted wave model
for $(\pi,K)$ inclusive spectra corresponding to $\Lambda$
and $\Sigma$ formation processes, using the Wigner transformation of the
nuclear density matrix.
The expression of the double differential cross section consists of
the incoming pion distorted wave function, the outgoing kaon distorted
wave function and the undetected hyperon distorted wave function
at each collision point, where the conservation of the classical
local momenta is respected. The momentum distribution of the bound nucleon
in the target nucleus is obtained from the Wigner transformation of
Hartree-Fock wave functions.

We have first applied the model to inclusive $(\pi^+ ,K^+)$ $\Lambda$
formation spectra on the $^{28}$Si and $^{12}$C targets measured at
KEK \cite{SAHA}. In this case, since the $\Lambda$ s.p. potential is
well known, there is no adjustable parameter. The standard $\Lambda$ potential
strength is found to reproduce well overall energy dependence of the data.
The strength is underestimated. However, this is a rather preferable result,
because the proper treatment of the CM motion has to be implemented
in our SCDW formulation and there should be some contributions from
two-step processes which are not taken into account. The quantitative
estimation of these effects is one of the important subjects to be investigated.

Observing from the $(\pi^+ ,K^+)$ $\Lambda$ formation spectra that
the SCDW model provides a useful description of the inclusive
spectrum without any adjustable parameters and renormalization
factors, we have proceeded to the $(\pi^- ,K^+)$ $\Sigma$ formation
spectra. The comparison of the calculated curves using several
choices of  the $\Sigma$ s.p. potential strength in a standard
Woods-Saxon geometry with
experimental data from KEK \cite{KEK,SAHA} has shown
that an attractive $\Sigma$-nucleus potential overestimates
the spectrum at lower excitation energies. Although there
are rather large uncertainties in the information of the
elementary process, we see that the repulsive potential is
necessary to account for the absolute strength of the spectrum.
Although we have to await quantitative estimation of various
effects above mentioned to specify the strength of the
$\Sigma$-nucleus potential, it is reasonable to conclude that
the $\Sigma$ hyperon experiences repulsion in nuclear medium
and its magnitude is not so strong as around 100 MeV which was
suggested by DWIA analyses in Ref. \cite{KEK}.

The information about the repulsive feature of the $\Sigma$-nucleus
potential constrains the two-body $\Sigma$-$N$ potential model and
thereby improves our understanding of the interactions between octet baryons.
In the literature there has been a few $\Sigma$-$N$ potential models which
predict repulsive $\Sigma$ mean field. In Nijmegen models \cite{NIJDF,NIJNS},
only the model F is satisfactory in this respect. Another model is a SU(6)
quark model by the Kyoto-Niigata group \cite{FU96a,FU96b,FU01}. The model
FSS gives 20 MeV \cite{KOH}, on the other hand the more sophisticated
version fss2 predicts somewhat smaller repulsion of 8 MeV. More studies
are certainly needed to determine the strength of the $\Sigma$ s.p.
potential, by employing various choice of the potential shapes.
The energy dependence of the $\Sigma$-nucleus potential may also
have to be taken into consideration.

The analyses of the data of heavier target nuclei are
important in the next step.
Since the neutron excess means that the $T=\frac{3}{2}$ contribution
becomes larger, we could check the isospin dependence
of the $\Sigma$-$N$ interaction on the basis of experimental
data.  The SCDW analyses of the data on $^{58}$Ni, $^{115}$In
and $^{209}$Bi taken at KEK \cite{SAHA} are in progress.

Finally we note that the present framework can be straightforwardly extended
to describe other inclusive spectra, such as $(K, \pi)$, $(K^-,K^+)$, $(\pi, \eta)$,
$(\gamma, K)$, $(\gamma, \eta)$ and so on.

\bigskip

This study is supported by Grants-in-Aid for Scientific
Research (C) from the Japan Society for the Promotion of
Science (Grant Nos. 15540284, 15540270 and 17540263).

\appendix
\section{Wigner transformation of the density matrix}
We present an explicit expression for the Wigner transformation of the
density matrix:
\begin{eqnarray}
 \Phi_h (\bR,\bK) &\equiv& \frac{1}{(2\pi)^3} \int d\bs \; e^{-i\bs\cdot\bK}
 \nonumber \\
 && \times \sum_h\;\phi_h^*(\bR-\frac{1}{2}\bs)\phi_h(\bR+\frac{1}{2}\bs).
\end{eqnarray}
We write the s.p. wave function of each partial wave in $r$-space as
\begin{equation}
\phi_h (\br) =\frac{1}{r} \phi_{n_h,\ell_h,j_h}(r)
 [ Y_{\ell_h}(\hat{\br})\times \chi_{1/2}]_{m_h}^{j_h},
\end{equation}
where $\phi_{n_h,\ell_h,j_h}(r)$ is a radial wave function and $\chi_{1/2}$
a spin part.
Let us denote the Fourier transform of the single-particle wave
function $\phi_h(\br)$ as $\tilde{\phi}_h(\bk)$.
\begin{eqnarray}
 \tilde{\phi}_h(\bk) &=& \frac{1}{(2\pi)^3} \int d\br
 e^{-i\bk\cdot\br} \phi_h (\br) \nonumber \\
 &=& \frac{1}{(2\pi)^{\frac{3}{2}}} i^{2n_h-\ell_h}
 [ Y_{\ell_h}(\hat{\bk})\times \chi_{1/2}]_{m_h}^{j_h} \nonumber \\
 & & \times \frac{1}{k} \tilde{\phi}_{n_h,\ell_h ,j_h}(k),
\end{eqnarray}
where $\hat{\bk}$ represents angular parts of the vector $\bk$ and
the Fourier transformation of the radial wave function is defined as
\begin{equation}
\frac{1}{k} \tilde{\phi}_{n_h,\ell_h,j_h}(k)
 =(-i)^{2n_h} \sqrt{\frac{2}{\pi}}
 \int dr\; rj_{\ell_h} (kr) \phi_{n_h,\ell_h , j_h}(r).
\end{equation}
$\Phi_h(\bR,\bK)$ in Eq. (A1) is rewritten as
\begin{equation}
 \Phi_h (\bR,\bK)=\int d\bp \;\tilde{\phi}_h^* (\bK-\frac{1}{2}\bp)
\tilde{\phi}_h (\bK+\frac{1}{2}\bp) \; e^{i\bp\cdot\bR}.
\end{equation}

Using the expression of Eq. (A3), we first obtain
\begin{eqnarray}
 & & \sum_{m_h} \tilde{\phi}_h^* (\bK-\frac{1}{2}\bp)
 \tilde{\phi}_h (\bK+\frac{1}{2}\bp) \nonumber \\
 &=& \frac{1}{(2\pi)^3}
 \sum_{m_h} \;[Y_{\ell_h}(\widehat{\bK-\frac{1}{2}\bp})\times
 \chi_{1/2}]_{m_h}^{j_h*} \nonumber \\
 & & \times  [Y_{\ell_h}(\widehat{\bK+\frac{1}{2}\bp})\times
 \chi_{1/2}]_{m_h}^{j_h}  \frac{1}{|\bK-\frac{1}{2}\bp |}
 \frac{1}{|\bK+\frac{1}{2}\bp |} \nonumber \\
 & & \times \tilde{\phi}_{n_h,\ell_h ,j_h}^*
 (|\bK-\frac{1}{2}\bp|) \tilde{\phi}_{n_h,\ell_h ,j_h}
 (|\bK+\frac{1}{2}\bp|).
\end{eqnarray}
Since the spin part gives $\chi_{1/2\:m_s}^* \chi_{1/2\:m_s'}
 \rightarrow \delta_{m_s m_s'}$, the recoupling of
angular momenta leads to the following expression.
\begin{eqnarray}
 & & \sum_{m_h} \tilde{\phi}_h^* (\bK-\frac{1}{2}\bp)
 \tilde{\phi}_h (\bK+\frac{1}{2}\bp) \nonumber\\
&=& \frac{1}{(2\pi)^3}
 \frac{2j_h+1}{4\pi} P_{\ell_h}
(\cos \theta_{\bK,\bp}) 
  \frac{1}{|\bK-\frac{1}{2}\bp |}
 \frac{1}{|\bK+\frac{1}{2}\bp |}  \nonumber \\
 & & \tilde{\phi}_{n_h,\ell_h ,j_h}^*
 (|\bK-\frac{1}{2}\bp|) \tilde{\phi}_{n_h,\ell_h ,j_h}
 (|\bK+\frac{1}{2}\bp|).
\end{eqnarray}
Here, the angle between $\bK-\frac{1}{2}\bp$ and $\bK+\frac{1}{2}\bp$
is denoted by $\theta_{\bK,\bp}$; that is,
\begin{equation}
 \cos \theta_{\bK,\bp} = \frac{\bK^2 -\frac{1}{4}\bp^2}
{|\bK-\frac{1}{2}\bp||\bK+\frac{1}{2}\bp|}.
\end{equation}
Then, Eq. (A5) becomes
\begin{eqnarray}
 \Phi_h (\bR,\bK) &=& \int d\bp \;e^{i\bR\cdot\bp}\; \frac{1}{(2\pi)^3}
\frac{2j_h+1}{4\pi} \nonumber \\
 & \times & P_{\ell_h}
(\cos \theta_{\bK,\bp}) \frac{1}{|\bK-\frac{1}{2}\bp |}
 \frac{1}{|\bK+\frac{1}{2}\bp |} \nonumber \\
& \times & \tilde{\phi}_{n_h,\ell_h ,j_h}^*
 (|\bK-\frac{1}{2}\bp|) \tilde{\phi}_{n_h,\ell_h ,j_h}
 (|\bK+\frac{1}{2}\bp|). \nonumber \\
\end{eqnarray}
Noting that the following relation holds
\begin{eqnarray}
 \int_0^{2\pi} d\phi_p \; e^{i\bR\cdot\bp} 
 &=& 2\pi \sum  i^\ell (2\ell +1) j_\ell (Rp) \nonumber\\
 &\times& P_\ell (\cos \widehat{\bK\bR}) P_\ell (\cos \widehat{\bK\bp}),
\end{eqnarray}
we obtain the Legendre expansion of $\Phi_h (\bR,\bK)$.
Each component $\Phi_h^\ell(R,K)$ of the expansion
\begin{equation}
\Phi_h (\bR,\bK)= \sum_{\ell=\mbox{even}} P_\ell (\cos \widehat{\bK\bR})
\; \Phi_h^\ell (R,K)
\end{equation}
is given as follows
\begin{widetext}
\begin{eqnarray}
 \Phi_h^\ell (R,K)&=& 2\pi (-1)^{\ell/2} (2\ell +1) \int p^2 dp
 d\cos \widehat{\bK\bp} \;j_\ell (Rp)P_\ell (\cos \widehat{\bK\bp})(2j_h +1)
 \frac{1}{(2\pi)^3}\frac{1}{4\pi} P_{\ell_h}(\cos \theta_{\bK,\bp}) \nonumber \\
 && \times \frac{2}{\pi} \int dr \; rj_{\ell_h}(|\bK+\frac{1}{2}\bp|r)
\phi_{n_h,\ell_h ,j_h}(r)  \int dr \; rj_{\ell_h}(|\bK-\frac{1}{2}\bp|r)
\phi_{n_h,\ell_h ,j_h}(r).
\end{eqnarray}
\end{widetext}

\end{document}